\documentclass[prl, amsfonts, twocolumn, nofootinbib, showpacs]{revtex4}
\usepackage{graphicx, epsfig}
\input epsf
\newcommand{\be}{\begin{equation}}
\newcommand{\ee}{\end{equation}}
\newcommand{\mpl}{M_{\rm Pl}}
\newcommand{\mbh}{M_{\rm BH}}
\newcommand{\etas}{\eta_{\rm s}}
\newcommand{\etaw}{\eta_{\rm w}}

\newcommand{\twd}{t_{\rm WD}}
\newcommand{\Twd}{T_{\rm WD}}
\newcommand{\tBH}{t_{\rm BH}}
\newcommand{\gapp}{\mathrel{\raise.3ex\hbox{$>$}\mkern-14mu
              \lower0.6ex\hbox{$\sim$}}}
\newcommand{\lapp}{\mathrel{\raise.3ex\hbox{$<$}\mkern-14mu
              \lower0.6ex\hbox{$\sim$}}}

\begin{document}

\title{Holes in the walls: primordial black holes as a solution to 
the cosmological domain wall problem}

\author{Dejan Stojkovic$^1$}
\author{Katherine Freese$^1$}
\author{Glenn D. Starkman$^{2}$}
\affiliation{$^1$MCTP, Department of Physics, University of Michigan,  Ann
Arbor, MI 48109-1120 USA}
\affiliation{$^2$ Department of Physics, Case Western Reserve University,
             Cleveland, OH~~44106-7079}

\begin{abstract}
\widetext
We propose a  scenario in which the cosmological
domain wall and monopole  problems are solved without any fine
tuning of the initial conditions or parameters in the Lagrangian
of an underlying filed theory. In this scenario domain walls sweep
out (unwind) the monopoles from the early universe, then the fast
primordial black holes perforate the domain walls, change their
topology and destroy them. We find further that the (old vacuum)
energy density released from the domain walls could alleviate but
not solve the cosmological flatness problem.
\end{abstract}

\pacs{???}
 \maketitle

Domain walls arise in a wide class of cosmological models.
Any early-universe phase transition, in which a discrete symmetry of
classical field theory is spontaneously broken,  results
in domain walls
%. In principle, classical field theory models involving
%spontaneous symmetries breaking, embedded
%in a particular cosmological model, which admit discrete symmetries
%accommodate the existence of the domain walls. Most of
%This includes the breaking of most grand unified theory (GUT)
% symmetries
\cite{ViSh}.
According to standard cosmology, domain walls, once formed,
quickly come to dominate the energy density of the universe and
severely violate many observational astrophysical
 constraints including measurements of the cosmic background radiation.
The domain walls must disappear before the epoch  of
nucleosynthesis at the latest.
Various solutions to this cosmological problem have been put forward,
notably a period of inflation \cite{inflation}, 
but it remains useful to examine new solutions.

Recently, two of us proposed a solution to the cosmological monopole
 problem:
primordial black holes, produced in the early
universe, can accrete magnetic monopoles within the horizon before
the relics dominate the energy density of the universe \cite{SF}.
Here we propose that primordial black holes can be used in yet
another way. We will show that primordial black holes can perforate
domain walls, and that the resulting holes in the walls can grow
to destroy the domain walls altogether.
We also note a variant of this picture in which monopoles
and domain walls can both be destroyed:
following the scenario presented in \cite{Tanmay,Brand},
it is possible for the domain walls to sweep up and eliminate
monopoles before the domain walls themselves are destroyed
(if they are sufficiently long-lived).

We imagine that domain walls are formed in the early universe,
during some phase transition at an energy scale $\eta$.
The energy per unit area of these domain walls is $\sigma \approx
\eta^3$. By the Kibble mechanism \cite{Kibble}, we expect to
form one domain wall per cosmological horizon at the formation time.
Subsequently, the cosmological
horizon grows to encompass many previously-disconnected
horizon volumes. The domain wall network evolves.
For walls arising from a remnant $Z_2$ symmetry, the network is usually
 dominated by one infinite
 wall of very complicated topology. In addition there are some finite
 closed walls. The structure can be more complicated for $Z_N$ and
 non-Abelian walls.  $Z_N$ (with large $N$)
 and non-Abelian walls  show a tendency for frustration
(freezing into a static structure). While $Z_2$ walls do not tend to
 frustration, their motion can still be severely damped if their
 interaction
 with surrounding matter is strong.
Thus, in most cases the network of domain walls is slowly evolving and
thus very slowly dissipates its energy.

As the scale factor $a(t)$ of the universe increases,
the number density of domain walls decreases as the volume,
$n \propto a^{-3}$,
but the mass of a slowly evolving domain wall increases approximately as
 its area, $m \propto a^{2}$.
The net effect is that the energy density of domain walls redshifts only
 as fast as $a^{-1}$.
Meanwhile, matter and radiation redshift as $a^{-3}$ and $a^{-4}$
 respectively.
Eventually (at least in the absence of  an even more slowly evolving
 component),
the domain walls come to dominate the energy density of the universe.

The characteristic time for wall domination is given by a generic
value $\twd = (G \sigma )^{-1}$. Since a domain-wall dominated
expansion has $a(t) \propto t^2$, rather than the $a(t)\propto
t^{1/2}$ of radiation domination, it would drastically change the
abundance of light elements produced during nucleosynthesis.
Therefore, if the universe produces domain walls, it must get rid
of them before nucleosynthesis.

The early universe may also produce large numbers of primordial black
holes. Such black holes can be formed by many processes
\cite{pbh1,pbh2,pbh3,pbh4,pbh5,pbh6,pbh7,pbh8,pbh9}. The earliest mechanism  
for black hole production can be fluctuations in the space-time metric at
the Planck epoch. Large number of primordial black holes can also be
produced by nonlinear density fluctuations due to oscillations of some
(scalar) field. If within some region of space density fluctuations are
large, so that the gravitational force overcomes the pressure, we can
expect the whole region to collapse and form a black hole. In the early
universe, generically, black holes of the horizon size are formed,
although it is possible to form much smaller black holes
\cite{pbh2}. Black holes can also be produced in first and second
order phase transitions in the early universe \cite{pbh5,pbh6}.
%Collision of bubbles
%of new vacuum in the first order phase transitions can be responsible for
%primordial black holes formation.
Gravitational collapse of cosmic string loops \cite{pbh7} and
closed domain walls can also yield black holes. The mass range of
primordial black holes formed in the above mentioned processes ranges
roughly from $\mpl$ (black holes formed at the Planck epoch) to $M_{\rm
sun}$ (black holes formed at the QCD phase transition). 

\begin{figure}[tbp]
\centerline{\epsfxsize = 0.9 \hsize \epsfbox{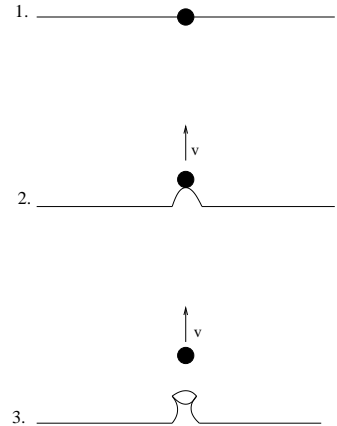}}
\caption{Possible outcomes of a domain wall-black hole encounter
 1)  the black hole gets stooped 2)
adiabatic interaction - the domain wall gets reconnected behind the
black hole 3) non-adiabatic interaction --- the black hole makes a hole
in the wall } \label{pierce} \end{figure}

As the black holes move through the universe, they encounter domain
walls.  The outcome of the encounter depends on the relative velocity
of the domain wall and the hole.
If the relative black hole-domain wall velocity is small, the
black hole gets stuck on the domain wall and its
kinetic energy goes into oscillatory modes  of the wall (configuration (1)
in Fig.~\ref{pierce}).  
If the black hole kinetic energy is large enough,
the black hole can pierce a hole in the wall as it passes through
(configuration (3) in Fig.~\ref{pierce}).
Depending on the underlying field  theory,
the boundary of a hole made in this way can be either a
cosmic string (if the field theory admits the existence of such strings)
or a black  string (a one-dimensional generalization of a black hole).
Even at intermediate velocities,  where one might have 
expected the wall to smoothly reconnect behind the black hole after
it passes through (configuration (2) in Fig.~\ref{pierce}),
the hole in the domain wall formed by its intersections with the black
 hole 
event horizon might instead begin to expand away from the black hole, 
bounded by a cosmic or black string.
This will happen if the intermediate  states  of the system 
are suitably unstable to such expansion.
 
Under certain generic conditions (shown below), the hole in the wall
 expands.
If there is only one hole in a large domain wall, it may not be
sufficient to entirely destroy the wall (which is expanding
itself).  However, if at least four
holes are made in the wall and expand outwards,
Chamblin and Eardley \cite{CE} have shown very generally  that
the domain wall will be utterly destroyed.  It is this mechanism
that we will employ to rid the universe of excess domain walls.

%\vspace{-.5cm}

The probability of cosmic/black string formation here
can be large due to thermal activation.  By contrast,
in situations where the process of cosmic/black string is spontaneous
(quantum mechanical tunneling or instanton), it can be highly
suppressed. The process we want to use here is induced by a black
hole and it is "over the barrier" process rather than tunneling.
The kinetic energy of the black holes gets converted into a thermal bath;
if this energy is large enough, it can cause the process of string
formation to be classically
allowed. On dimensional grounds we expect the
probability for this process to be proportional to
$e^{-m_s/T}$, where $m_s$ is the mass of the created string,
while $T$ is the temperature of the thermal bath. When $T \sim m_s$, the
process becomes unsuppressed. For an effective thermal bath one needs
finite energy distributed over some finite volume relevant for the process
--- a condition that is rarely satisfied in collisions of point like
particles. Since both the black hole and the domain wall are extended
objects we expect that most of the change in kinetic energy of the black
hole in a black hole-wall collision is converted 
into a thermal bath. For the exact
fraction of the kinetic energy that goes into thermal bath, one would have
to perform a detailed analytical (or numerical) analysis.

We have mentioned that is possible for
black strings to be formed at the black hole/domain 
wall encounter.
Domain walls tend to decay by nucleating string loops on their
world-sheet. Even topologically stable domain walls can be unstable to
nucleating black strings when we include gravity. The situation
here has an analog in one dimension less. It is well known that a
topological string can break by nucleating a pair of black holes.
Unfortunately, there is no appropriate calculation for the case of
a hole bounded by a black string within a domain wall sheet. Here,
we assume that such a configuration is allowed. It would be very
interesting to find an appropriate metric that describes this
configuration.

The probability for  black string creation should also be greatly
enhanced (in comparison with spontaneous black string nucleation), not
only due to thermal bath enhancement but also due to the physical setup. A
black string solution is just a direct product of a black hole and one
extra dimension. There are known solutions of $(3+1)$-dimensional black
strings in string theory where there are some additional fields beside
gravity. Alternatively, we can construct a black string solution as a
direct product of a BZT black hole (the only known black hole
solution in $(2+1)$ dimensions) and an extra spatial dimension.
There are no asymptotically flat (real) black hole solutions in
$(2+1)$ dimensions. The BZT black hole is a solution in the
presence of the cosmological constant. The momentum energy tensor
of the domain wall is such that the interior of the wall is a
region with negative pressure due to the (old vacuum) energy
density that acts similarly to a cosmological constant. One can easily
imagine that in the black hole-domain wall encounter, the black
hole horizon gets deformed and stretched by the domain wall
tension into a black string configuration on the boundary of a
hole made in the wall. We leave detailed description of this
process for the future.

A comprehensive study of the interaction of the domain walls and
black holes in the weak field regime and/or some other approximations has
been done in \cite{FSS}. A proper treatment of the present problem in the
strong field regime, near the black hole horizon, is merited; however, in
the absence of such a treatment, we will rely on simple
back-of-the-envelope estimates.

We first consider the kinematics of a black hole perforating a
domain wall. This is a classically allowed process if the black
hole has sufficient kinetic energy to cause the production of a
(black or cosmic) string bounding  the perforation. In the
non-relativistic limit this is crudely: \be \mbh(1 + v^2/2) + \pi
R_p^2 \sigma \geq \mbh' (1  + v'^2/2) + 2\pi R_p \mu
\label{energyconservation} \ee where $\mbh$ and $\mbh'$ are the
mass of the black hole before and after the interaction and $v$
and $v'$ its velocity. $R_p$ is the initial radius of the
perforation in the wall, $\sigma$ is the energy per unit area of
the domain wall, and $\mu$ is the energy per unit length of the
string bounding the
 perforation.
Generically, $\sigma \sim \etaw^3$ where $\etaw$ is the energy scale at
 which the walls are formed.
Similarly, $\mu \sim \etas^2$ where $\etas$ is the energy scale
 characteristic of the strings.
(For definiteness  we consider the situation where domain walls can be
 bounded by a cosmic string. The calculation with black strings is
 analogous
with $\mu_s$ replaced by $\mu_{bs}$ --- mass per unit length of
the black string.)

One can take for $R_p$ the Schwarzschild radius of the
black hole after the interaction. In the interaction, a black hole
swallows a piece of the domain wall.
 Accounting for the negative pressure
inside the wall is difficult in our heuristic calculation of energy
conservation. Consequently we confine ourselves to the case where the
energy  of the eaten disc of domain wall is small compared to the black
hole mass, 
\be
\label{eq:disc}
\mbh \gg \pi R_p^2 \sigma. \ee
In this case we can still take the
radius of the perforation in the domain wall to be the Schwarzschild
radius of the black hole after the interaction, $R_p = 2\mbh'/\mpl^2$.
The mass of the black hole is enhanced by an amount equal
to the mass of the eaten disk,
 \be
\label{eq:mbhprime}
\mbh' = \mbh + \pi R_p^2 \sigma \sim \mbh
\left(1 + 4\pi \frac{\etaw^3 \mbh}{\mpl^4}\right)
\label{MBHprime} \ee
to first non-trivial order in small quantities.

Then the condition in Eq. (\ref{eq:disc}) is satisfied as long as

\be\label{Mup}
\mbh < {1 \over 16 \pi} \mpl \left({\mpl \over \eta_w} \right)^3 . \ee
For $\etaw \sim 10^{17}$GeV (motivated by GUTs), the mass must be
\be
\mbh \leq 2 \times  10^4\mpl \left({\eta_w \over 10^{17}{\rm GeV}}
\right)^{-3} .
\ee
We note that, at these very early stages of the universe,
such small black hole masses are extremely plausible (the horizon
size of the universe at the time was extremely small).

We can now estimate $v'$, the post-collision velocity of the black hole
 (in the frame of the domain wall) by assuming that all the momentum   of
the incoming black hole is carried  off by the outgoing black hole,
and none is transferred to the domain wall.  (We expect this too to be a
good approximation if (\ref{Mup})  holds.)
From momentum conservation $\mbh' v' = \mbh v$ and using
Eq. (\ref{eq:mbhprime}), we have
\be v' \sim v \left(1 + 4\pi \frac{\etaw^3
\mbh}{\mpl^4}\right)^{-1} . \ee

Under these assumptions, the condition (\ref{energyconservation}) can be
 rewritten more usefully:
\be
2 \frac{4\pi \mbh \etaw^3}{\mpl^4} \geq \frac{\etas^2 \mpl^2}{\etaw^3
 \mbh} -
\frac{v^2}{2} .
\label{KElimit_1}
\ee
This is algebraically satisfied if either
\be
\frac{\mbh}{\mpl} \geq \frac{1}{\sqrt{8\pi}} \frac{\etas \mpl^2}{\etaw^3}
\label{KElimita}
\ee
or
\be
\frac{1}{2} \mbh v^2  \geq  \frac{\etas^2 \mpl^2}{\etaw^3} \, .
\label{KElimitb}
\ee
We will concentrate our analysis on these two conditions.

It may be worth noting that perforation of the wall may also occur
by Hawking radiation of a closed string from the black hole 
as it traverses the domain wall. 
This is possible only if the mass of the string loop does not greatly
exceed the Hawking temperature of the black hole.  Assuming as
above that condition (\ref{eq:disc}) is satisfied, 
this means 
\be \frac{\mbh}{\mpl} \geq \frac{1}{\sqrt{8 \pi}} \frac{\mpl}{\etas}  .
 \ee 
Of course, this is not a sufficient condition -- 
the Hawking radiation of an extended coherent state such as a string loop
could be highly suppressed. On the other hand, Hawking radiation
of ordinary particles is an s-wave process, so the extended coherent
 nature
of the string loop should not in and of itself be regarded as
problematic, especially if a significant portion of released energy
from Hawking radiation goes into the thermal bath.  Nevertheless, 
we henceforth confine our attention to the classical perforation of the
 string.

The tension of the string  bounding the perforation tends to close
the hole in the domain wall, while the domain wall tension tends to
stretch the hole. The hole continues to expand  if \be 0 \leq
\frac{\partial}{\partial r}\left(2 \pi r \mu - \pi r^2 \sigma
 \right)\vert_{R_p}
= 2 \pi \left(\etas^2 - \frac{2 \mbh}{\mpl^2} \etaw^3 \right)  .
\label{expandcondition}
\ee
(To this order the gravitational potential energy of the string in the
 field of the black hole
is $r$-independent.)
Reorganizing this equation, the hole expands if
\be \label{Mlow}
\frac{\mbh}{\mpl} \geq \frac{1}{2}\left( \frac{\etas}{\mpl}
 \right)^2\left(
 \frac{\mpl}{\etaw} \right)^3 .
\ee This is a weaker constraint than (\ref{KElimita}) and also
weaker than (\ref{KElimitb}) so long as $\etas \leq
\frac{1}{\sqrt{2\pi}} \mpl$ (which we shall take to be the case),
so a classically induced  perforation will expand.

The reader may be concerned that we have not yet 
ensured that the black-hole-domain-wall  collision does not end
with the wall reconnecting behind the black hole.
Condition (\ref{KElimita}) or (\ref{KElimitb})  ensures
that there is enough energy to form a string. 
Condition (\ref{expandcondition}) ensures that a string, 
if formed, grows.  
Therefore, if these conditions are met, 
then string loop formation and expansion is energetically favored.
It is therefore  likely to occur in at least a reasonable fraction
of hole-wall collisions.

It is useful to write the condition (\ref{Mup}) (for validity of
our approximation) together with the conditions for formation of a
hole in the wall (\ref{KElimita}) and (\ref{KElimitb}) as

% \be \frac{1}{\sqrt{8\pi}}\frac{\etas
%\mpl^2}{\etaw^3}  \leq \frac{\mbh}{\mpl}  \leq  \frac{ \mpl^3}{16 \pi
%\etaw^3} \, .\ee

\be \label{min} {\rm min}\left(\frac{1}{\sqrt{8\pi}}\frac{\etas
\mpl^2}{\etaw^3}, \frac{2}{v^2}\frac{\etas^2 \mpl}{\etaw^3}\right)
  \leq \frac{\mbh}{\mpl}  \leq  \frac{
\mpl^3}{16 \pi \etaw^3} \, .\ee

For $\etas \sim \etaw \sim 10^{17}$GeV,  this is 
\be \label{102}
{\rm min}\left( 2\times 10^{3}, \frac{2 \times 10^{2}}{v^2}
\right) \leq \frac{\mbh}{\mpl} \leq
 2 \times 10^4  .
\ee
This condition (\ref{min}) is independent of $\etaw$ and can be satisfied
for the right black hole mass.
This is a strong indication that there are no serious problems
for the scenario to work. Now, we put the scenario into a cosmological
framework.

The  observational constraints on the abundance of primordial black
holes have been studied earlier \cite{pbh8}. These constraints are
sensitive to the assumptions of the model but for the range of
parameters of interest for GUT domain walls, observations do not
practically imply any constraints (unless the endpoint of black
hole evaporation is a Planck mass relic).

The maximal mass of a primordial black hole is limited by the total mass
 within the
cosmological horizon, i.e. $M_{\rm hor} = \mpl^3/\Lambda^2$  at any given
 energy scale $\Lambda$ at which the black hole forms.  This is also the
 expected mass scale of a
black hole in most early universe scenarios  for the production of black
 holes. (Stellar black holes are a clear counterexample from the present
 universe.) Taking
\be \label{bhmass}
\mbh = f \mpl^3/\Lambda^2 , \ee
where $f$ is the fraction
of the horizon mass that equates to the mass of a black hole,
equation (\ref{Mup}) becomes

\be \frac{\Lambda}{\mpl} >  \sqrt{16\pi f}
 \left(\frac{\etaw}{\mpl}\right)^{3/2} .
\label{selfconsistent_b}
\ee

We remind the reader that perforation of the domain wall is
kinematically allowed if either Eq. (\ref{KElimita}) or
(\ref{KElimitb}) is satisfied. Let us examine each of these in turn.
Satisfying the condition  (\ref{KElimita}) would require that
\be
\label{cond1}
\frac{\Lambda}{\mpl}  \leq \sqrt{\sqrt{8\pi} f}
 \left(\frac{\mpl}{\etas}\right)^{1/2}
\left(\frac{\etaw}{\mpl}\right)^{3/2} . \label{KElimita_b} \ee
The combination of Eqs. (\ref{selfconsistent_b}) and
(\ref{KElimita_b}) can be satisfied if
$\etas \leq \frac{1}{\sqrt{32 \pi}}\mpl$.

The other possibility for satisfying the kinematics of the collision
is given by Eq. (\ref{KElimitb}). We now examine the consequences
of this condition.
The kinetic energy $\mbh v^2 /2 $ of a black hole of mass  $\mbh$
formed at energy scale $\Lambda$ would be expected to be of order
 $\Lambda$,
\be \frac{1}{2} \mbh v^2   \sim \Lambda  . \ee
 For $\mbh = f
\mpl^3/\Lambda^2$, this means that we expect \be v_{\rm thermal}
\sim  \sqrt{\frac{2}{f}} \left(
 \frac{\Lambda}{\mpl}\right)^{3/2}  .
\label{vthermal} \ee This is quite small for $f\sim1$ if $\Lambda
\ll \mpl$; however, if the mass of the black hole is much
smaller than the horizon mass ($f<<1$) its velocity can be relativistic.
(Black holes formed by the gravitational collapse of cosmic string
loops would be expected to be relativistic.) To obtain estimates
we will take the black hole's velocity at its formation $v_i$ to be as
given by (\ref{vthermal}) with $f=1$. Imposing the constraint
(\ref{KElimitb}) would then require

\be \label{vcondition}
\frac{\Lambda }{\mpl} \geq \left(\frac{\etas}{\mpl}\right)^2
 \left(\frac{\mpl}{\etaw}\right)^3
\ee
Requiring $\Lambda \leq \mpl$, we then have

\be \label{etasupper} \frac{\etas}{\mpl} \leq
\left(\frac{\etaw}{\mpl}\right)^{3/2} , \ee a rather mild hierarchy
if $\etaw\sim 10^{17}$GeV.

The black holes may form either before the domain walls ($\Lambda \geq
 \etaw$),
or after them ($\Lambda \leq \etaw$).  If they form after the walls,
there  remain two possibilities -- the black holes can form
either before or after
the walls come to dominate the energy density  of the
universe.  Black holes form before wall domination if
$\mpl/\Lambda^2 \sim t_\Lambda \leq \twd \sim (G \sigma)^{-1}$, i.e., if
\be
\label{BHbeforeWD}
\frac{\Lambda}{\mpl} \geq
 \left(\frac{\etaw}{\mpl}\right)^{3/2}
\ee

If black holes form before wall domination, then, as long as they don't
 decay,
they characteristically travel a distance
$d_\leq = \int_{t_\Lambda}^{t_{WD}} v dt$.
Taking $v_i$ to be  the initial thermal velocity of the black  hole
given by (\ref{vthermal}), and using the fact that velocity
scales as $a^{-1}$ during the expansion of the universe and that
the universe is radiation dominated
before the wall domination, we have $v = v_i {a_\Lambda \over a}
=v_i ({t_\Lambda \over t})^{1/2}$. Then

\be d_\leq = d(t\leq\twd) \sim
2 v_i \sqrt{t_\Lambda \twd} \, . \ee
%before wall domination
%and an additional
%\be
%d_\geq  = d(t\geq\twd) \sim  v_i \sqrt{t_\Lambda \twd}
%\ee
%after wall domination (assuming that the walls are not  annihilated).

The
number of black holes formed at time $t_\Lambda$ that are
inside a horizon volume at a later time $\twd$ is

\be \label{countBHs} N_{\rm BH} = {\rm F} \left(
\frac{\twd}{\frac{\Lambda}{\Twd} t_\Lambda}
 \right)^3
\, ,
\ee
where $\Twd=\sqrt{\mpl/\twd} = \left(\etaw/\mpl\right)^{3/2}\mpl$ is the
 temperature
at time $\twd$, and $F$ is the number of black holes formed per horizon
 volume at time $t_\Lambda$.
If black holes are formed by density perturbations, we can expect
to have one black hole per horizon at the time of formation.
In general, $F$ may be greater or less than $1$, although $Ff \leq 1$.
$N_{\rm BH}$ is easily evaluated:
\be
\label{countBHs_b}
N_{\rm BH} = {\rm F} \left(\frac{\Lambda}{\mpl}\right)^3
 \left(\frac{\mpl}{\etaw}\right)^{9/2} .
\ee

The number of perforations which one expects  in a domain wall by
the time $\twd$ is therefore

\be \label{Nperforations} N_{\rm perforations} \sim
\frac{d_\leq}{\twd} N_{\rm BH} \approx \frac{ F}{{\sqrt f}}
\left(\frac{\Lambda}{\mpl}\right)^{7/2}
\left(\frac{\mpl}{\etaw}\right)^3  . \ee

Since most of the energy density is concentrated in one infinite wall
of complicated geometry/topology, it is sufficient to destroy only
this one.  The other finite closed pieces of walls will collapse
within some short time due to their own tension.  We remind the reader
that as long as there are at least four black hole perforations into
the domain wall, the wall can generically be destroyed \cite{CE}.
With generic values $F \sim 1$ and $f \sim 1$, the condition $N_{\rm
  perforations} \geq 4 \sim {\cal O}(1)$ becomes

\be \label{enoughperforations}
\frac{\Lambda}{\mpl} \geq  \left(\frac{\etaw}{\mpl}\right)^{6/7}  .
\ee
This condition is stronger than the one in (\ref{BHbeforeWD}),
which is required for black holes to form before domain walls
come to dominate (we have assumed this latter condition in deriving
Eq. (\ref{enoughperforations})).

Again, we now split our discussion according to the two possible
conditions in Eqs.(\ref{KElimita}) and (\ref{KElimitb}) for satisfying
the kinematics required for a black hole to perforate a domain wall.
If we are to satisfy condition (\ref{cond1}) (required by
Eq. (\ref{KElimita})) together with (\ref{enoughperforations}) we must
have

\be \label{Lcon1}
 \left(\frac{\mpl}{\etas}\right)^{1/2}
\left(\frac{\etaw}{\mpl}\right)^{3/2}  \geq \frac{\Lambda}{\mpl} \geq
\left(\frac{\etaw}{\mpl}\right)^{6/7} \, , \ee
where we have dropped factors of order unity.
The consistency condition (the upper limit is greater than the lower) is

\be \label{con01}
\frac{\etas}{\mpl} \leq  \left(\frac{\etaw}{\mpl}\right)^{9/7}  \, .
\ee
Condition (\ref{con01}) is the necessary condition for the solution of
the domain wall problem, under the assumption of Eq. (\ref{KElimita})
for satisfying the kinematic requirements.

We now turn to the second possible kinematic requirement of
Eq. (\ref{KElimitb}). Then we need to satisfy (\ref{vcondition}) together with
(\ref{enoughperforations}), so that

\be \label{max1}
\left(\frac{\Lambda}{\mpl}\right) \geq {\rm max} \left[
\left(\frac{\etaw}{\mpl}\right)^{6/7}, \
\left(\frac{\etas}{\mpl}\right)^{2}\left(\frac{\mpl}{\etaw}\right)^{3}
\right] . \ee

 This equation, together with the constraint on
 $\etas$ given in Eq. (\ref{etasupper},) provides the set of conditions
for the  solution of the domain wall problem given the second possibility
in  Eq. (\ref{KElimitb}) for satisfying the kinematic requirements for
 perforation.

 The black hole lifetime ($\tBH \approx \mbh^3/\mpl^4$) is very short.
For  example, a black hole of $\mbh = 10^5\mpl$ evaporates within about
 $10^{-13}$s.  Therefore, just making the required black holes is not
 enough; we must also ensure that they do not evaporate before they
 perforate the domain walls.  The above calculation assumed that the
 black holes did not decay before the wall domination era, i.e.,

\be \tBH = \mbh^3/\mpl^4  \geq  \frac{\mpl^{2}}{\etaw^{3}} = \twd \, . \ee
Using Eq. (\ref{bhmass}), this equation becomes

\be
\frac{\Lambda}{\mpl} \leq \left( \frac{f \etaw}{\mpl}
\right)^{1/2}  \, . \label{BHnotdecay}
\ee

Again we split our discussion into the two ways the kinematics of wall
perforation can be satisfied.  The above equation can be consistent with
(\ref{BHbeforeWD}) and (\ref{enoughperforations}) if $f \geq
(\etaw/\mpl)^{5/7}$, which is satisfied if $f\sim 1$ as expected. Thus,
from (\ref{Lcon1}) and (\ref{BHnotdecay}) we have (for $f\sim 1$)

\be \label{nc12}
{\rm min} \left[ \left(\frac{\mpl}{\etas}\right)^{1\over 2}
\left(\frac{\etaw}{\mpl}\right)^{3 \over 2} ,
\left( \frac{\etaw}{\mpl} \right)^{1 \over 2} \right]
\geq \frac{\Lambda}{\mpl} \geq
\left( \frac{ \etaw}{\mpl} \right)^{6/7} \, .\ee
This condition takes care that black holes are capable of perforating the
domain wall and do not decay before the wall domination. The constraint on
$\etas$ that goes together with this one is given in (\ref{con01}).

If, alternatively, we assume the second condition in Eq. (\ref{KElimitb})
for the kinematics,
the alternative set of necessary conditions  can be obtained from
(\ref{max1}) and (\ref{BHnotdecay})

\be \label{Lmax1}
 \left( \frac{\etaw}{\mpl}
\right)^{1/2}  \! \geq  \! \left(\frac{\Lambda}{\mpl}\right) \geq {\rm max}
\left[ \left(\frac{\etaw}{\mpl}\right)^{6/7}, \!
\left(\frac{\etas}{\mpl}\right)^{2} \! \left(\frac{\mpl}{\etaw}\right)^{3}
\right]\, . \ee
The constraint on $\etas$ that goes together with the above equation
is given in (\ref{etasupper}).

The black holes that evaporate before wall domination are still
capable of destroying walls. In this case, the condition (\ref{nc12})
could be somewhat relaxed.  Alternatively, it is also possible that
black holes form only after wall domination. However, in this case it
becomes more and more difficult for the black holes to solve the
domain wall problem.  Since in the wall dominated universe $a(t) \sim
t^2$, black hole velocities will be redshifted very quickly; in
addition the walls rapidly move away from one another. These two
effects make the solution of the domain wall problem unlikely if the black 
holes form only after wall domination.

In summary, there are two alternative sets of conditions for the
solution of the domain wall problem, depending on which condition we
use to satisfy the required kinematics for a black hole to perforate a
domain wall. If either set of conditions is satisfied, domain walls
disappear.  The first set is determined by the condition
(\ref{KElimita}). For this set, the necessary condition for the
solution of the domain wall problem is given by Eq. (\ref{con01}) if
the black holes are formed at energy scales given in Eq. (\ref{nc12}).
The second set is determined by the condition (\ref{KElimitb}). For
this set, the necessary condition for the solution of the domain wall
problem is given by Eq. (\ref{etasupper}) for the black holes that are
formed at energy scales given in Eq. (\ref{Lmax1}).  It is enough to
satisfy either one of these sets of conditions in order for our
scenario to work.  Thus, combining the righthand side of
Eq. (\ref{nc12}) (which is less restrictive than that of
Eq. (\ref{Lmax1})) with the left hand side of Eq. (\ref{Lmax1}) (which is
less restrictive than the left hand side of Eq. (\ref{nc12})), we
conclude that the black holes that are formed at energy scales

\be \label{bh} \left(\frac{\etaw}{\mpl}\right)^{1/2} \geq
\frac{\Lambda}{\mpl} \geq \left(\frac{\etaw}{\mpl}\right)^{6/7} \, ,
\ee
are generically capable of destroying walls, as long there is the
following hierarchy between the black/cosmic string scale and domain
wall scale:

\be  \label{string} \frac{\etas}{\mpl} \leq
\left(\frac{\etaw}{\mpl}\right)^{9/7} \, . \ee

Here, the right hand side of the equation is obtained from
Eq. (\ref{con01}) (which is less restrictive than the right hand side
of Eq. (\ref{etasupper})) and there is no lower bound (since there is
none in Eq. (\ref{etasupper})).  It is not difficult to envision
satisfying these bounds for GUT domain walls with $\etaw \sim (10^{16}
-10^{17})$GeV.

Outside of the range of parameters 
given by eq. (\ref{bh}) and (\ref{string}), it 
is still possible that the walls get destroyed
under some special circumstances.
For example, it is quite likely for a primordial black hole
to form near the domain wall; in this case,
the black hole does not have to travel an entire horizon distance 
to encounter the wall, so that the probability of wall destruction
is significantly enhanced.  Alternatively, if a black hole is formed with 
(or gets accelerated to)  a velocity much larger than the average thermal 
velocity, then again the probability of wall destruction is enhanced. 
Either of these two scenarios is in fact quite plausible.

We have seen that, within the range of parameters given by eq. (\ref{bh}) 
and (\ref{string}), domain walls have little chance to survive. 
While there are several constraints  on this mechanism
for  black hole destruction of domain walls, it seems to work
quite generically for GUT-scale domain walls. The specific
constraints can easily be recalculated for domain walls formed by
a phase transition at any other energy scale.

{\it Solving the monopole problem:}~~ In order to complete the scenario we
mention an interesting result presented in \cite{Tanmay,Brand}. Namely,
it was shown that the interaction between the monopoles and domain walls
can be such that monopoles unwind while passing through the domain wall
and their topological charge gets spread all over the wall. Thus, the
domain walls can sweep up monopoles from the universe. The basic concept
of the idea as well as details about the interaction can be found in
\cite{Tanmay,Brand}.  For the reasonable range of the parameters in the
model, the domain walls
clean the universe of monopoles before the critical time $\twd$ when
the walls come to dominate the universe. This naturally fits into our
scenario.

{\it Solving other cosmological problems?:}~~
The equation of state of domain walls
depends on their  velocity. If they are relativistic,
then the equation of state is $p=1/3 \rho_w$, as for ordinary radiation.
But, if their velocity is small, the equation of state is $p = - 2/3
 \rho_w$ \cite{ViSh}.  This result is not surprising since the pressure
inside the domain wall is negative. This is consistent with the scaling $
\rho_w \sim 1/a$. Since this is a slower redshifting that even the
curvature term in the  Friedmann equation,
it suggests that a period of wall domination in the early universe could
 probably solve the flatness problem.
If the energy density of the universe is dominated by the walls energy
density the scale factor grows quadratically with
time, i.e. $a \sim t^2$, so what we have here is nothing more than
domain-wall-driven power law inflation.
During the expansion of the hole
in a domain wall, the vacuum energy contained in the wall gets released.
This process recalls post-inflationary re-heating.

However, a rough analysis indicates that reheating after
domain wall inflation would not work.
After the period of inflation needed for solving the cosmological flatness
and horizon problems, the domain walls end up very far apart. The energy
released when the domain walls disappear is at first located only near the
domain walls. Even if this energy travels away from the walls at the
speed of light, thermalization of a big enough patch to
encompass our observable universe takes too long.
If we require that the thermalization is finished
at some high enough temperature (say $T >$MeV), then the amount of
inflation would not be sufficient to completely solve the flatness and
horizon problems.

For completeness we note here that the mechanism of destroying domain walls 
by primordial black holes can also fit naturally in models where cosmologically dangerous domain walls appear after inflation (see for example brane inflation models \cite{brane_inflation}).

We would like to conclude with the summary of the  cosmological
scenario we proposed here:

1. Domain walls sweep out (unwind) the monopoles

2. Black holes perforate the domain walls and destroy them

3. Energy  density released from the domain walls could alleviate
but not solve the cosmological flatness and homogeneity problems.

\vspace{12pt} {\bf Acknowledgments}:\ \
The authors are grateful to Valeri Frolov, Tanmay Vachaspati and Nemanja
Kaloper for very useful conversations. DS is grateful to Valeri Frolov
with whom a similar problem was discussed in a slightly different
framework of brane world models. DS and KF acknowledge support
from the DOE and the Michigan Center for Theoretical Physics at the
University of Michigan.

{9}

\end{document}